\documentclass[twocolumn,showpacs,preprintnumbers,amsmath,amssymb]{revtex4}
\usepackage{amsmath,amssymb,graphics,epsfig,subfigure}
\usepackage{color}

\begin{document}

\thispagestyle{empty}

\begin{center}

\title{Gravitational equal-area law and critical phenomena of cuspy black hole shadow}

\date{\today}
\author{Shao-Wen Wei$^{a,b}$,
Chao-Hui Wang$^{a,b}$,
Yu-Peng Zhang$^{a,b}$,
Yu-Xiao Liu$^{a,b}$,
Robert B. Mann$^{c}$}

\affiliation{$^{a}$Lanzhou Center for Theoretical Physics, Key Laboratory of Theoretical Physics of Gansu Province,
Key Laboratory of Quantum Theory and Applications of MoE,
Gansu Provincial Research Center for Basic Disciplines of Quantum Physics, Lanzhou University, Lanzhou 730000, China,\\
$^{b}$Institute of Theoretical Physics $\&$ Research Center of Gravitation,
School of Physical Science and Technology, Lanzhou University, Lanzhou 730000, China,\\
$^{c}$Department of Physics $\&$ Astronomy, University of Waterloo, Waterloo, Ont. Canada N2L 3G1}

\begin{abstract}
The formation of a cusp on a black hole shadow is a striking signature of physics beyond the Kerr paradigm. Using the Konoplya-Zhidenko  metric, a general parametrized deformation of Kerr black hole, we demonstrate that this morphological change fundamentally alters the shadow's topology with the topological charge flipping from 1 to -1. To analyze this topological transition, we introduce a gravitational equal-area law, analogous to Maxwell's construction in thermodynamics, and identify a critical point for cusp formation. Near this point, we uncover universal behavior characterized by a critical exponent 1/2, which places this gravitational lensing system within the mean-field universality class. These results establish a new framework for testing fundamental physics of black hole shadows, reframing the search for deviations from general relativity as a targeted hunt for a distinct topological and critical phenomenon.
\end{abstract}


\maketitle
\end{center}

\section{Introduction}

The black hole shadow is a direct manifestation of spacetime curvature in the strong-field regime, providing a powerful probe for testing general relativity \cite{Synge,Luminet}. Its size and shape, determined primarily by the black hole's mass and spin, offer a clean window into the fundamental nature of gravity \cite{Bardeen,Chandrasekhar,Hioki}.

The landmark Event Horizon Telescope observations of M87* and Sagittarius A* brought this theoretical prediction into an observational reality, confirming the predictions of general relativity and constraining the mass and spin of these supermassive objects \cite{EHT,EHTb}. Crucially, these images also serve as a stringent test of theoretical frameworks: any deviation from the predicted Kerr shadow would be conclusive evidence for new physics, such as modified gravity, exotic compact objects, or surrounding dark matter \cite{EHTc,Gralla,Herdeiro,Perlick}. A particularly compelling approach to testing the Kerr hypothesis is through parametrized deformations of the Kerr metric, which introduce a small number of free parameters that can be constrained by shadow observations in a model-agnostic fashion, analogous to the parametrized post-Newtonian (PPN) formalism.

Deviations from the Kerr metric, often parameterized as ``hair", can dramatically alter a black hole's shadow. The presence of matter fields like scalar or axion hair can deform the standard D-shape shadow into more complex patterns \cite{Cunha,Kuang,Huang}. A particularly intriguing feature is the formation of a cusp, a sharp point on the shadow boundary. These cusps signify stable fundamental photon orbits or spherical orbits, potentially linked to novel spacetime instabilities \cite{Cunhap,Sanchis}. For instance, Ref. \cite{Cunhap} confirms the emergence of stable photon orbits or light ring induced instabilities via fully nonlinear numerical simulations of ultracompact bosonic stars that are free of all other established instabilities. These instabilities drive the stars to either migrate toward nonultracompact configurations or collapse into black holes. Furthermore, Ref. \cite{Sanchis} shows that the backreaction of energy trapped in the vicinity of stable light rings can induce fundamental modifications to the spacetime geometry. It is worth mentioning that this cuspy phenomenon emerges in various non-Kerr scenarios, including black holes with Proca hair or those embedded in dark matter halos \cite{Carlos,Wang,Qian,Ohta}.

While cuspy shadow morphologies have been identified in a handful of specific black hole spacetime models, the universal physical principle governing the formation of such cuspy structures has thus far remained elusive. In this paper, we uncover the fundamental underlying principles that dictate the emergence of cuspy black hole shadows, leveraging two robust and mutually complementary theoretical frameworks. First, through a topological analysis, we demonstrate that the presence of a cusp fundamentally alters the shadow's global topology. The Kerr shadow possesses a topological charge of +1, whereas the cuspy shadow has a charge of -1, placing them in unequivocally distinct topological classes. Second, we reveal a deep analogy between cusp formation and thermodynamic phase transitions. The appearance of the cusp is a gravitational analogue of swallowtail behavior in the free energy, exhibiting universal critical behavior and admitting a gravitational Maxwell equal area-law construction to precisely determine the self-intersection points of the observable shadow boundary. Although we illustrate our results using the Konoplya-Zhidenko (KZ) metric, the three central findings, the topological charge flip, the equal-area law, and the critical exponent $1/2$, are universal across all non-Kerr spacetimes that support stable fundamental photon orbits.

The manuscript is structured as follows. In Sec. \ref{cbhs}, we review the Cuspy black hole shadow by taking the KZ black hole as an example. In Sec. \ref{tcc}, we consider the topology of the black hole shadow. The equal area law is constructed in Sec. \ref{geal}. Meanwhile, the critical phenomena are explored in Sec. \ref{cpp}. Finally, in Sec. \ref{Conclusion}, we summarize and discuss our results.

\section{Cuspy black hole shadow}
\label{cbhs}

To explore the formation of cusps within an observationally relevant framework, we adopt the rotating KZ black hole \cite{Konoplya}. The KZ metric is a parametrized deformation of Kerr: it introduces a single additional parameter $\eta$ (with dimensions of $M^3$) that quantifies the deviation from general relativity in the near-horizon region. When $\eta = 0$, the metric reduces exactly to Kerr. Nonzero $\eta$ can be interpreted as encoding, in a model-agnostic way, the aggregate effect of any beyond-GR physics that modifies the near-horizon geometry, such as additional matter fields, quantum gravity corrections, or environmental effects. This parametrized approach mirrors the philosophy of the PPN formalism: one does not need to commit to a specific alternative theory of gravity to test GR; one simply measures the deformation parameter and checks whether it is consistent with zero.

In the Boyer-Lindquist coordinates, the line element for this deformed black hole solution reads \cite{Konoplya}
\begin{eqnarray}
	\label{xy}
	ds^{2}=& -&\bigg(1-\frac{2Mr^2+\eta}{r\rho^{2}}\bigg)dt^{2}
	+\frac{\rho^{2}}{\Delta}dr^{2}+\rho^{2} d\theta^{2}\nonumber\\
	&+&\sin^{2}\theta\bigg[r^{2}+a^{2}
	+\frac{(2Mr^{2}+\eta)a^2\sin^{2}\theta}{r\rho^{2}}\bigg]d\phi^{2}\nonumber\\
	&&-\frac{2(2Mr^2+\eta)a\sin^{2}\theta}{r\rho^{2}}dtd\phi,
\end{eqnarray}
where the metric functions $\Delta$ and $\rho^{2}$ are given by
\begin{equation}
	\Delta=a^{2}+r^{2}-2Mr-\frac{\eta}{r},\quad \rho^{2}=r^{2}+a^{2}\cos^{2}\theta.
\end{equation}
The constants $M$, $a$, and $\eta$ correspond, respectively, to the black hole mass, spin, and deformation parameter. The event horizon radius $r_h$ is determined by solving the condition $\Delta(r_h)=0$. This construction alters gravitational dynamics exclusively in the near-horizon vicinity while preserving asymptotic flatness at large radii. It thus constitutes a tractable theoretical framework for probing deviations from the Kerr paradigm within the near-horizon spacetime regime. To further characterize the geometry, we derive the effective energy–momentum tensor sourcing the modified metric. Restricting attention to the equatorial plane $\theta=\pi/2$, the nonvanishing tensor components take the forms
\begin{eqnarray}
 T_{tt}&=&\frac{\eta  \left(-5 a^2 r+2 \eta +4 M r^2-2r^3\right)}{r^8},\\
 T_{rr}&=&\frac{2 \eta }{r^2 \left(a^2 r-\eta -2 M r^2+r^3\right)},\\
 T_{\theta\theta}&=&-\frac{3 \eta }{r^3},\\
 T_{\phi\phi}&=&\frac{\eta  \left(-5 a^4 r+2 a^2 \left(\eta +2 M r^2-4
 	r^3\right)-3 r^5\right)}{r^8},\\
 T_{t\phi}&=&T_{\phi t}=\frac{a \eta  \left(5 a^2 r-2 \eta -4 M r^2+5
 	r^3\right)}{r^8}.
\end{eqnarray}
In the limit $\eta\to 0$, the solution recovers the standard Kerr black hole geometry, for which the effective energy-momentum tensor identically vanishes, $T_{\mu\nu}=0$. At the event horizon $r=r_h$, the radial component $T_{rr}$ exhibits a divergence, whereas all remaining tensor components remain finite. We additionally compute the effective energy density measured by an observer with tour-velocity $n^{\mu}=(1, 0, 0, 0)$, defined via the contraction
\begin{eqnarray}
 \rho&=&T_{\mu\nu}n^{\mu}n^{\nu}\nonumber\\
  &=&\frac{\eta  \left(-5 a^4 r+a^2 \left(3 \eta +6 M r^2-7
 	r^3\right)-2 r^5\right)}{r^5 \left(a^2 \left(\eta +2 M
 	r^2+r^3\right)+r^5\right)}.
\end{eqnarray}
Evaluating this density at the horizon $r=r_h$ yields the simplified expression
\begin{eqnarray}
 \rho=-\frac{2\eta}{r_{h}^5}.
\end{eqnarray}
This relation immediately shows that positive deformation parameters $\eta>0$ generate negative effective energy density, while negative $\eta<0$ produces positive energy density

Notably, all components of $T_{\mu\nu}$ and the effective density $\rho$ decay to zero asymptotically as $r\to\infty$. This asymptotic behavior demonstrates that the geometric deformations induced by $\eta$ can confined primarily to near-horizon region. Accordingly, negative and positive values of $\eta$ mimic the gravitational behavior of ordinary baryonic matter and phantom matter, respectively.

From an observational perspective, the KZ metric's key virtue is that its deformation parameter $\eta$ is directly accessible to measurement. EHT observations of M87 and Sgr A* have already placed constraints on parametrized non-Kerr metrics, including the KZ family \cite{EHTc}, by comparing the observed shadow size and shape to theoretical predictions. The cusp transition we identify introduces a qualitatively new observable: rather than measuring subtle continuous shape deformations, one can ask the binary question of whether the shadow lies in the smooth phase ($\eta > \eta_c$) or the cuspy phase ($\eta < \eta_c$). A detection of cuspy features would therefore constitute a discrete, unambiguous falsification of the Kerr hypothesis.

The KZ spacetime also naturally hosts radially stable spherical photon orbits when $|\eta|$ is sufficiently large — a geometric feature that is the necessary condition for cusp formation. This same feature appears in physically well-motivated non-Kerr scenarios, including black holes with Proca hair and those embedded in dark matter halos \cite{Carlos,Wang,Qian,Ohta}.

Although the distinctive morphological feature of a cuspy black hole shadow was presented in Ref. \cite{Wang}, the intrinsic physical mechanisms responsible for this peculiar shadow structure remain unaddressed and have not been systematically investigated in existing studies. In this work, we adopt the KZ spacetime as an observationally motivated, analytically tractable representative to demonstrate that a gravitational equal-area law and associated critical phenomena emerge whenever stable spherical orbits are present. Notably, this constitutes a universal result that extends beyond the context of the KZ black hole, indicating broader applicability across a class of non-Kerr black hole spacetimes.

The shadow boundary is determined by unstable spherical photon orbits. Owing to the system's integrability, these orbits can be directly projected onto the celestial coordinates ($\alpha$, $\beta$) of a distant observe \cite{Bardeen}
\begin{eqnarray}
 \alpha=-r_0\frac{p^{\phi}}{p^{r}},\quad
 \beta=r_0\frac{p^{\theta}}{p^{r}},
\end{eqnarray}
where $p^{\mu}$ measures the projection of the four momentum of photons moving in the black hole background. For a distant observer ($r_0\to\infty$) in the equatorial plane, these coordinates are given by \cite{Wang}:
\begin{eqnarray}
\alpha&=&\frac{6Mr^4-2r^5+5\eta r^2+a^2(\eta-2r^2(M+r))}{2a(M-r)r^2-a\eta},\label{alpha}\\
\beta&=&\pm\frac{r^2\sqrt{8a^2(2Mr^3+3r\eta)-(6Mr^2-2r^3+5\eta)^2}}{a(2r^2(r-M)+\eta)} \label{beta}
\end{eqnarray}
where $r$ is the radius of the fundamental photon orbits. The deformation parameter $\eta$ breaks the Kerr degeneracy, thus allowing for super-radiant spin values $(a>M)$ even while an event horizon is preserved.

\begin{figure}
\includegraphics[width=4.5cm]{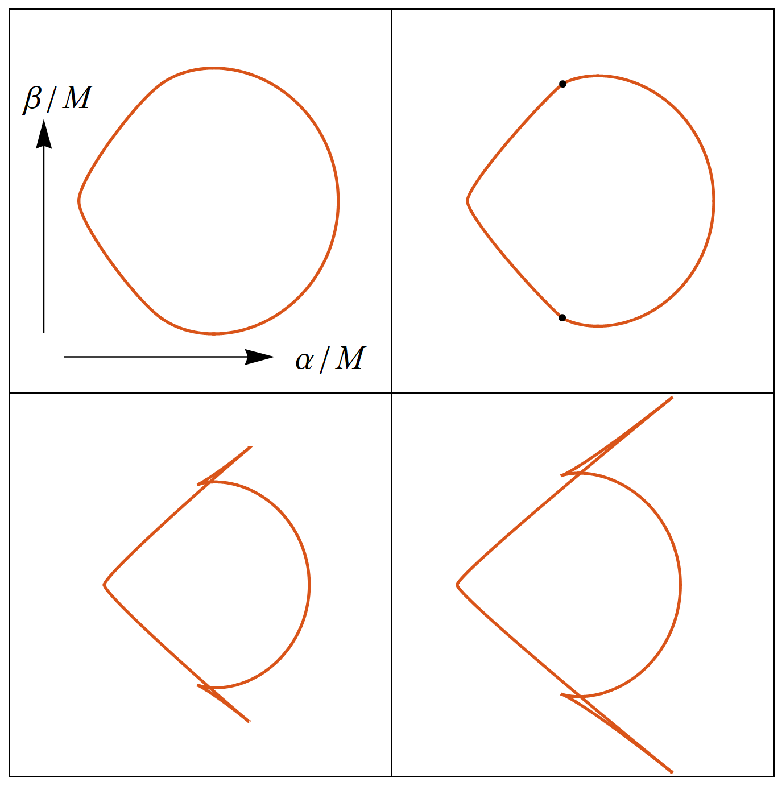}
\caption{Shadows cast by the KZ black hole with $a/M=2$. The deformation parameter takes values of $\eta/M^3=$1.5, 1.052, 0.6, and 0.5. Black dots denote the critical points where cusps are about to form. The bottom panels clearly show the resulting cuspy shadow structures.}\label{pShadow}
\end{figure}

We fix the spin at $a/M=2$ and use the deformation $\eta$ to drive the system through a topological transition. The resulting shadow morphology is shown in Fig. \ref{pShadow}. For large values of $\eta$, the shadow is a simple closed curve, distinct from the Kerr case yet topologically equivalent. As $\eta$ is lowered to a critical value of $\eta_c \approx 1.052 M^3$, a pair of critical points (cusps) marked by black dots, appear on the shadow boundary. Below this threshold ($\eta < \eta_c$), the boundary develops self-intersections, forming two distinct cuspy regions that grow as $\eta$ decreases. This qualitative transition from a simple closed curve to a self-intersecting one constitutes the central phenomenon studied in this paper, which we will analyze through the frameworks of topology and critical phenomena.

\section{Topological charge}
\label{tcc}

Beyond its visual appearance, the global structure of a black hole shadow can be characterized by a topological charge \cite{Wei}. For standard black holes, this charge is a robust invariant ($\delta=1$), distinguishing them from objects like naked singularities which possess a different topology. Here, we demonstrate that the formation of a cusp fundamentally alters this invariant, thereby driving a genuine topological phase transition.

The topological charge is defined by the winding number of a tangential vector around the shadow boundary via the Gauss-Bonnet theorem \cite{Wei}:
\begin{eqnarray}
\delta=\frac{1}{2\pi}\left(\oint\frac{dl}{\mathcal{R}}+\sum_i\Delta\theta_i\right).\label{jif}
\end{eqnarray}
The first term measures the integral over the smooth parts of the boundary, where $\mathcal{R}$ is the local radius of curvature. The second term accounts for a discrete sum over exterior angles $\Delta\theta_i$ at all non-differentiable cusp points.

\begin{figure}
\includegraphics[width=7cm]{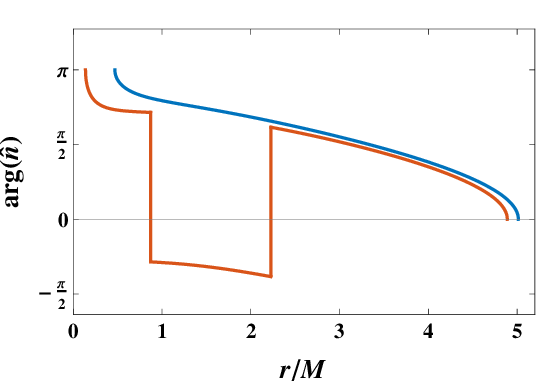}
\caption{Argument of the normal vector $\hat{n}$ along the shadow boundary parameterized by $r/M$ for $a/M=2$. The deformation parameter $\eta/M^3=$1.5 and 0.5 are described by the top blue and bottom red curves, respectively. Note that the curve is segmented for $\eta/M^3=$0.5, and both the differences of the sudden change in the argument are $\Delta\theta=\pi$. As $\eta/M^3$ decreases, the length of the lower segment of the red curve increases.}\label{pArgn}
\end{figure}

For a smooth, non-cuspy shadow ($\eta > \eta_c$), the boundary is differentiable everywhere, and the second term vanishes. The integral can be evaluated by considering the slope of the boundary
\begin{eqnarray}
 \mathcal{F}=\frac{d\beta}{d\alpha},\label{fff}
\end{eqnarray}
which is an important quantity and helpful in constructing the Maxwell equal law in the next section. By changing the integration variable to the radius parameter $r$ of the fundamental photon orbits and leveraging the shadow's $\mathcal{Z}_2$ symmetry, the charge simplifies to
\begin{eqnarray}\label{jifz2}
 \delta=\frac{1}{\pi} \left(\arctan \mathcal{F}(r_{min})-\arctan \mathcal{F}(r_{max})\right),
\end{eqnarray}
where $r_{min}$ and $r_{max}$ correspond to the leftmost and rightmost points of the shadow, as indicated in Fig. \ref{pShadow}. Considering that $\mathcal{F}(r_{max})=-\infty$ and $\mathcal{F}(r_{min})=+\infty$, one easily arrives the result
\begin{eqnarray}
 \delta=1.
\end{eqnarray}
This confirms that the non-Kerr shadow, so long as it remains smooth, belongs to the same topological class as the Kerr shadow. This can also be obtained by the argument of the normal vector $\hat{n}$ along the shadow boundary parameterized by $r$. Taking $\eta/M^3=1.5$ as an example (top blue curve in Fig. \ref{pArgn}), where no cusp patterns are observed, it is evident that as $r$ ranges from $r_{max}$ to $r_{min}$, the argument progressively increases from 0 to $\pi$, yielding $\delta=\pi/\pi=1$.

However, the topology changes dramatically in the presence of cusps (e.g. $\eta/M^3=0.5$). As shown in Fig. \ref{pArgn}, the cusps introduce sharp, discontinuous jumps in the argument of the boundary's normal vector described by the bottom red curve. Each cusp is associated with an angle jump of $\pi$. One might expect the angle jumps contributed by these two cusps to cancel each other. However, the normal vector $\hat{n}$ undergoes a clockwise rotation at both cusps, such that each cusp individually contributes $\Delta\theta =-\pi$. Including these contributions and exploiting the $\mathcal{Z}_2$ symmetry the topological charge \eqref{jifz2}
for the cuspy shadow becomes
\begin{eqnarray}
\delta=\frac{1}{\pi}(\pi+\Delta\theta_1+\Delta\theta_2) = \frac{1}{\pi}(\pi - \pi - \pi) = -1.
\end{eqnarray}
This result is profound: the topological charge flips from $\delta=1$ to $\delta=-1$. This is not a small perturbation but a fundamental change in the global properties of the shadow. It rigorously proves that shadows with and without cusps belong to entirely different topological classes, with the cusp acting as topological genus.

\section{Gravitational equal-area law}
\label{geal}

The self-intersecting boundary of a cuspy shadow is strikingly reminiscent of the swallow tail behavior of the free energy for a van der Waals fluid undergoing a liquid-gas phase transition. In thermodynamics, the unphysical, multi-valued region of the isotherm is resolved by the Maxwell equal-area law, which precisely determines the coexistence pressure. Here, we construct an analogous law to identify the self-intersection point of the shadow.

\begin{figure}
\subfigure[]{\label{Efa5}\includegraphics[width=4.2cm]{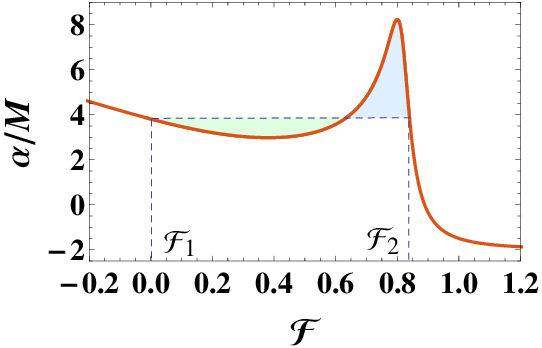}}
\subfigure[]{\label{Efb6}\includegraphics[width=4.2cm]{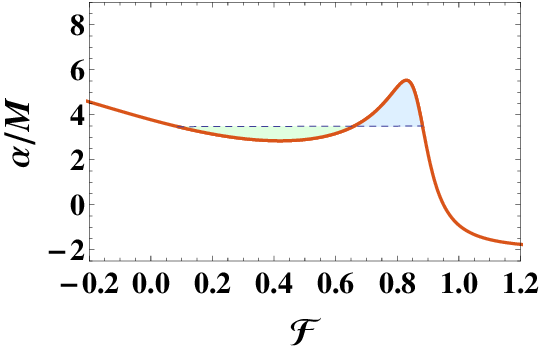}}
\subfigure[]{\label{Efa5}\includegraphics[width=4.2cm]{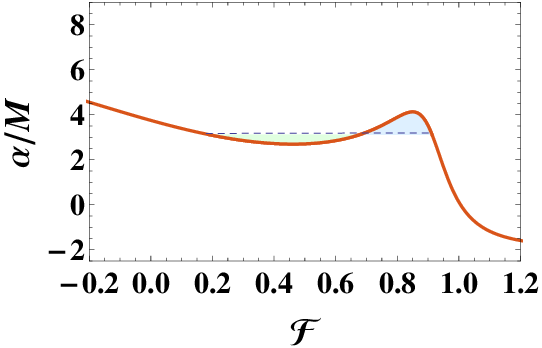}}
\subfigure[]{\label{Efb6}\includegraphics[width=4.2cm]{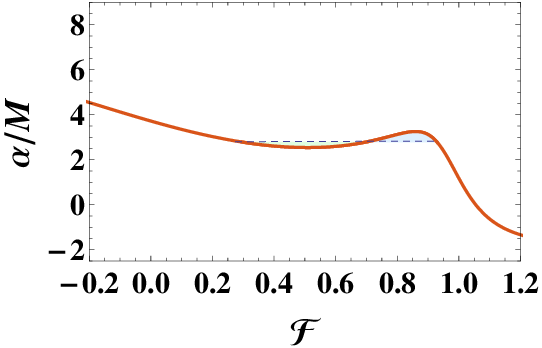}}
\caption{Gravitational equal-area law for the cuspy black hole shadow with $a/M=2$. (a) $\eta/M^3$=0.5. The two equal areas are each 3.316,  with the slopes $\mathcal{F}_1$=-0.0169 and $\mathcal{F}_2$=0.8397. (b) $\eta/M^3$=0.6. The two equal areas are 2.755. (c) $\eta/M^3$=0.7. The two equal areas are 0.134. (d) $\eta/M^3$=0.8. The two equal areas are 0.064. The horizontal dashed lines are for $\alpha_*/M$=3.872, 3.469, 3.109, and 2.785 for (a)-(d), respectively.}\label{pEfb6}
\end{figure}

We can treat the shadow boundary as a thermodynamic-like curve in the ($\mathcal{F}$, $\alpha$) plane, where $\mathcal{F}$, as defined in Eq. (\ref{fff}), serves as the variable conjugate to the celestial coordinate $\alpha$. For a fixed black hole configuration (with constant $a$ and $\eta$), the closed loop integral on the shadow's state space must vanish. At a self-intersection point, two different photon orbits (parameterized by $r_1$ and $r_2$) coincide at the same celestial coordinates ($\alpha_*, \beta_*$). The condition that the loop integral $\oint d\beta$ vanishes along a loop connecting these two points leads directly to our gravitational equal-area law:
\begin{eqnarray}
\oint \mathcal{F}d\alpha=0,
\end{eqnarray}
or equivalently
\begin{eqnarray}
 \int_{\mathcal{F}_1}^{\mathcal{F}_2}\alpha d\mathcal{F}=\alpha_*(\mathcal{F}_2-\mathcal{F}_1),
\end{eqnarray}
where $\mathcal{F}_2$ and $\mathcal{F}_1$ correspond to the same $\alpha_*$, a relationship linked to the self-intersection point that is under consideration for determination. This dictates that the self-intersection point $\alpha_*$ must be chosen such that it equalizes the two areas enclosed by the non-monotonic curve and a vertical line at $\alpha_*$ in the ($\mathcal{F}, \alpha$) plane, as illustrated in Fig. \ref{pEfb6}.

We numerically verify this law for the KZ black hole with $a/M=2$. As illustrated in Fig. \ref{pEfb6}, constant-$\eta$ curves below the critical value ($\eta < \eta_c$) exhibit the characteristic non-monotonic behaviour required for a first-order-like phase transition. By applying our equal-area construction, we can precisely locate the self-intersection point. For $\eta/M^3=0.5$, the law uniquely identifies the intersection at $\alpha_* \approx 3.872$ with conjugate variable values $\mathcal{F}_1=-0.0168743$ and $\mathcal{F}_2=0.839669$. This corresponds to two distinct fundamental photon orbits with radii $r_1/M \approx 3.164$ and $r_2/M \approx 0.642$, respectively. These values are in perfect agreement with direct numerical results, confirming the validity of our construction.

As $\eta$ increases towards the critical value $\eta_c$, the enclosed area (see $\eta/M^3=$ 0.6, 0.7, and 0.8 in Fig. \ref{pEfb6}) gradually decreases and vanishes exactly at the critical point. This behavior is perfectly analogous to a thermodynamic system near its critical point. This establishes the equal-area law as a robust and fundamental tool for analyzing the morphology of non-Kerr black hole shadows. Finally, in Sec. \ref{Conclusion}, we summarize and discuss our results.

\section{Critical phenomena}
\label{cpp}

\begin{figure}
\subfigure[]{\label{Expalpha}\includegraphics[width=4cm]{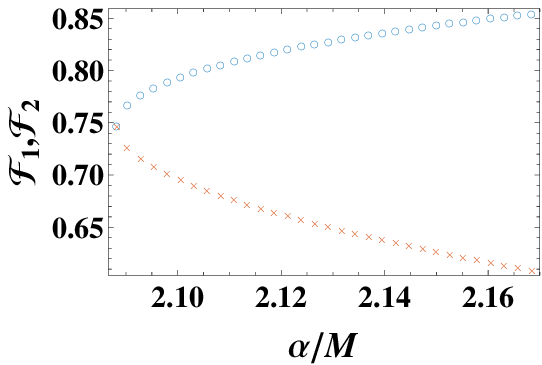}}
\subfigure[]{\label{Expeta}\includegraphics[width=4cm]{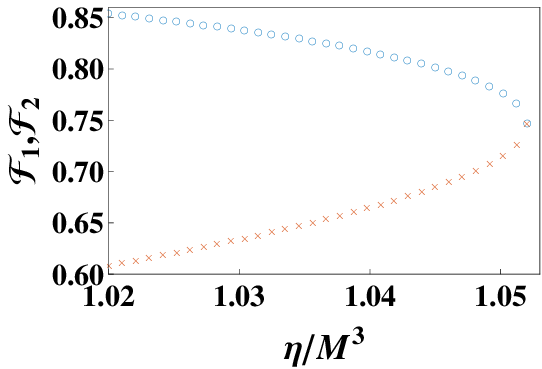}}
\caption{Numerical results of $\mathcal{F}_1$ (bottom curves) and $\mathcal{F}_2$ (top curves) with $a/M=2$ near the critical point. The critical value of $\mathcal{F}$ is 0.747. (a) $\mathcal{F}$ vs. $\alpha/M$. The critical point is at $\alpha_c/M$=2.088.  (b) $\mathcal{F}$ vs. $\eta/M^3$. The critical point is at $\eta_c/M^3$=1.052.}\label{pF12b}
\end{figure}

The transition from a smooth to a cuspy shadow is a critical phenomenon characterized by a non-analytic change. The critical point---the exact threshold where the cusps first appear---occurs when the two areas in our equal-area construction vanish. This corresponds to an inflection point in the ($\alpha, \mathcal{F}$) plane, defined by the conditions:
\begin{eqnarray}
 \left(\frac{\partial \alpha}{\partial \mathcal{F}}\right)_{a,\eta}=0,\quad \left(\frac{\partial^2 \alpha}{\partial \mathcal{F}^2}\right)_{a,\eta}=0.
\end{eqnarray}
Taking $a/M=2$, we identify the critical point as $\alpha_c/M=2.088$ and $\eta_c/M^3=1.052$. Instances where $\alpha>\alpha_c$ or $\eta<\eta_c$ will manifest the cuspy shadow. By employing the equal-area law illustrated in Fig. \ref{pEfb6}, we can derive the respective values of $\mathcal{F}_1$ and $\mathcal{F}_2$. These numerical outcomes are depicted in Fig. \ref{pF12b}. As $\alpha$ and $\eta$ approach their critical values, $\mathcal{F}_1$ increases while $\mathcal{F}_2$ decreases, indicating a reduction in the cuspy patterns. At the critical point, $\mathcal{F}_2=\mathcal{F}_1$, and the cuspy patterns converge towards a point, beyond which the pattern completely disappears.

\begin{figure}
\subfigure[]{\label{Expalpha}\includegraphics[width=4cm]{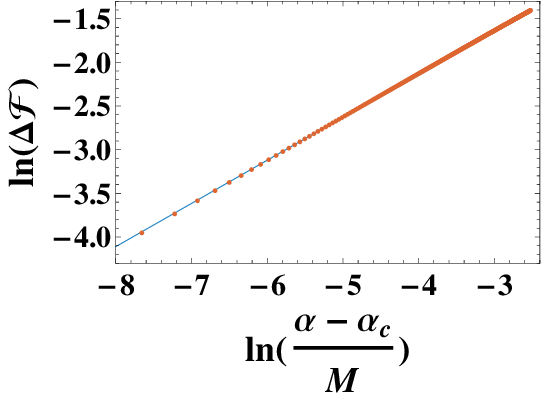}}
\subfigure[]{\label{Expeta}\includegraphics[width=4cm]{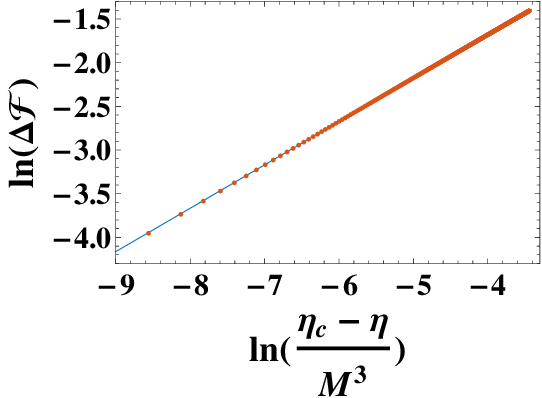}}
\caption{Critical behavior of $\Delta\mathcal{F}=\mathcal{F}_2-\mathcal{F}_1$ with $a/M=2$. Dots represent numerical results, and blue lines correspond to fitting curves. (a) $\Delta\mathcal{F}$ vs. $\ln(\frac{\alpha-\alpha_c}{M})$. The slope of the blue line is 0.495. (b) $\Delta\mathcal{F}$ vs. $\ln(\frac{\eta_c-\eta}{M^3})$. The slope of the blue line is 0.496. }\label{pExpalpha}
\end{figure}

Near a critical point, physical systems often exhibit universal behavior characterized by critical exponents.  Analogous with thermodynamic phase transitions \cite{Kubiznak,Weib}, we treat the difference between the self-intersecting branches, $\Delta\mathcal{F}=\mathcal{F}_2-\mathcal{F}_1$, as the order parameter. This order parameter should scale as a power law with the distance from the critical point:
\begin{eqnarray}
 \Delta\mathcal{F}\sim\left(\frac{\alpha-\alpha_c}{M}\right)^{\zeta_1},\\
 \Delta\mathcal{F}\sim\left(\frac{\eta_c-\eta}{M^3}\right)^{\zeta_2}
\end{eqnarray}
with $\zeta_i$ defined as the respective critical exponents. To test this hypothesis, we numerically computed $\Delta\mathcal{F}$ as a function of ($\alpha-\alpha_c$) or ($\eta_c-\eta$) using our equal-area law. The results are shown on a log-log plot in Fig. \ref{pExpalpha}. These data points fall perfectly on straight lines, and a linear fit yields  $\zeta_1=0.495$ and $\zeta_2=0.496$. This alignment strongly supports the conclusion that the critical exponents are $\zeta_1=\zeta_2=\frac{1}{2}$.

We further confirm this by analyzing the behavior of the corresponding unstable fundamental photon orbits, $r_1$ and $r_2$. The difference $\Delta r=r_2-r_1$ also serves as an excellent order parameter. Power-law fits of $\Delta r$ versus ($\alpha-\alpha_c$) and ($\eta_c-\eta$) yield critical exponents of 0.4999 and 0.5018, respectively.

Both calculations robustly yield a critical exponent of
\begin{eqnarray}
\zeta = \frac{1}{2}.
\end{eqnarray}
This result is highly significant. A critical exponent of 1/2 is the hallmark of the mean-field universality class (where the critical exponent $\beta=1/2$), which describes a vast range of physical systems, from the Curie-Weiss model of magnetism to the van der Waals liquid-gas transition. Our results establish that  cusp formation in a black hole shadow is not merely analogous to a thermodynamic phase transition---it is a direct manifestation of the same fundamental universality class.

\section{Summary}
\label{Conclusion}

We have established a new framework for understanding the shadows of non-Kerr black holes, revealing that   cusp formation is not a simple morphological change but a true topological phase transition. By connecting strong-field gravitational lensing to the physics of topology and thermodynamical critical phenomena, we uncovered the fundamental principles governing these exotic shadow shapes.

Our investigation yielded three key results. First, we proved that cuspy shadows are topologically distinct from their smooth counterparts, characterized by a topological charge of $\delta=-1$ in stark contrast to the $\delta=1$ of Kerr-like shadows. Second, we introduced a novel gravitational equal-area law, analogous to Maxwell's construction in thermodynamics, which provides a robust and predictive tool for analyzing the shadow's structure. Finally, by treating the onset of cusp formation as a critical phenomenon, we identified a universal critical exponent of $\zeta=1/2$. This finding rigorously places this gravitational system into the mean-field universality class, establishing a profound link between black hole optics and statistical mechanics.

Our results establish promising new directions for theoretical investigations and observation. Spacetimes hosting black holes with purely unstable fundamental photon orbits display conventional shadow morphologies, while cuspy shadow structures arise when stable fundamental photon orbits are geometrically allowed. Accordingly, the emergence of a cuspy shadow provides a direct observational diagnostic for confirming the existence of stable fundamental photon orbits. Moreover, our study no longer presents the KZ geometry itself as an established astrophysical model. Instead, it uses this analytically tractable example to expose a general local mechanism that may be tested in physically derived non-Kerr solutions, such as these black holes endowed with Proca hair, embedded within dark matter halos, and other compact black objects \cite{Carlos,Wang,Qian,Ohta}. 

The critical behavior uncovered in this work indicates that near-threshold black holes, those with $|\eta| \approx \eta_c$, may host distinctive observational signatures, including abrupt luminosity variations as the system fluctuates across the cusp transition. The KZ deformation parameter $\eta$ can be constrained both through shadow morphology (EHT and next-generation VLBI) and through gravitational-wave ringdown spectroscopy (LIGO-Virgo-KAGRA, LISA, and the Einstein Telescope). The topological transition we have identified converts these continuous constraints into a binary test: if future observations consistently yield smooth shadows with $\delta = +1$, the Kerr paradigm is corroborated; the detection of a single cuspy shadow with $\delta = -1$ would constitute a discrete falsification. In addition, the correspondence between stable fundamental photon orbits and accretion flow dynamics suggests such topological imprints can manifest as distinguishable structural features within accretion disks. Our framework provides the necessary tools to model these effects, transforming the search for new physics from a hunt for subtle deviations into a targeted search for the distinct topological and critical phenomena revealed in this work.

{\emph{Acknowledgements}.}---This work was supported by the National Natural Science Foundation of China (Grants No. 12475055, No. 12475056, No. 12247101) and by the Natural Science and Engineering Research Council of Canada.

\end{document}